\documentclass[prb,twocolumn,superscriptaddress,showpacs,amsmath,amssymb]{revtex4}
\usepackage{amsfonts}
\usepackage{bbm}
\usepackage{graphicx}
\usepackage{dcolumn}
\usepackage{bm}
\usepackage{subfigure}
\usepackage{color}

\begin{document}
\title{Perfect crossed Andreev reflection in the proximitized graphene/superconductor/proximitized graphene junctions}

\author{Shu-Chang Zhao}
\author{Lu Gao}
\affiliation{School of Science, Qingdao University of Technology, Qingdao, Shandong 266520, China}
\author{Qiang Cheng}
\email[]{chengqiang07@mails.ucas.ac.cn}
\affiliation{School of Science, Qingdao University of Technology, Qingdao, Shandong 266520, China}
\affiliation{International Center for Quantum Materials, School of Physics, Peking University, Beijing 100871, China}

\author{Qing-Feng Sun}
\email[]{sunqf@pku.edu.cn}
\affiliation{International Center for Quantum Materials, School of Physics, Peking University, Beijing 100871, China}
\affiliation{Hefei National Laboratory, Hefei 230088, China}
\affiliation{CAS Center for Excellence in Topological Quantum Computation, University of Chinese Academy of Sciences, Beijing 100190, China}

\begin{abstract}
We study the crossed Andreev reflection and the nonlocal transport in the proximitized graphene/supercondcutor/proximitized graphene junctions with the pseudospin staggered potential and the intrinsic spin-orbit coupling.
The crossed Andreev reflection with the local Andreev reflection and the elastic cotunneling being completely eliminated can be realized for the electrons with the specific spin-valley index when the intrinsic spin-orbit couplings in the left graphene and the right graphene possess the opposite sign.
The perfect crossed Andreev reflection with its probability equal to $100\%$ can be obtained in the space consisting of the incident angle and the energy of the  electrons.
The crossed conductance and its oscillation with the superconductor length are also investigated. The energy ranges for the crossed Andreev reflection without the local Andreev reflection and the elastic cotunneling are clarified for the different magnitudes of the pseudospin potential and the spin-orbit coupling.
The spin-valley index of the electrons responsible for the perfect crossed Andreev reflection can be switched by changing the sign of the intrinsic spin-orbit coupling or exchanging the biases applied on the left graphene and the right graphene. Our results are helpful for designing the flexible and high-efficiency Cooper pair splitter based on the spin-valley degree of freedom.
\end{abstract}
\maketitle

\section{\label{sec1}Introduction}
Cooper pair splitting can produce the spatially separate spin-entangled electrons in the normal conductor/superconductor/normal conductor junctions, which not only can help test the Bell's inequalities in experiment but also possesses applications in the fields of quantum information and quantum communication\cite{Hofstetter1,Chtchelkatchev,Sauret,Nilsson,Veldhorst,Hofstetter2,Tan,Pandey}. As the inverse process of the Cooper pair splitting, the crossed Andreev reflection (CAR) can convert the incoming electron from one normal conductor to the hole in the other conductor\cite{Recher,Zhu}. The strength of CAR is a direct measure of the efficiency of the Cooper pair splitting. The realization of CAR has attracted many theoretical and experimental researches\cite{Beckmann,Yamashita,Melin1,Melin2,Zimansky,Benjamin,Chen84,Zhang,Wang,YLiu,Zhang2,Reeg,Lu1,Liu,Galambos,Gul}.
For example, the influence of the even-odd interference behavior related to the Majorana quasiparticle on CAR is clarified in the junctions including the topological superconductor nanowire\cite{Liu}.
The CAR between the spin-polarized chiral edge states is obtained due to the Meissner effect in the quantum Hall regime of the two-dimensional electron gas\cite{Galambos}.
The possible signature for CAR across the superconductor separating two fractional quantum Hall states is observed at the particlelike fillings\cite{Gul}. Recently, the CAR and the spin cross-correlation in the quantum dot-superconductor-quantum dot devices are directly measured in experiment\cite{GZWang,Bordoloi}.

However, the emergence of CAR is usually accompanied by the local Andreev reflection (LAR), the normal reflection, and the elastic cotunneling (ECT) processes.
The LAR process will convert the incoming electron from one conductor
into the hole in the same conductor,
while the ECT (normal reflection) process will directly
transmit (reflect) the incoming electron from one conductor
to the other (same) conductor.
The presence of the LAR will weaken the occurrence and spoil the observation of the Cooper pair splitting since its inverse processes do not contribute
to the nonlocal entangle of electrons.
Many proposals have been proposed to eliminate LAR and ECT and
to realize the perfect CAR\cite{Zhou,YTZhang,SBZhang,Jakobsen,Lu2,Fuchs,Li,Reinthaler}.
For example, the perfect CAR can be realized through tuning the phase of the wave function for the Majorana fermion to be $\pi$ in the system consisting of the quantum anomalous Hall insulator coupled with a topological superconductor\cite{Zhou,YTZhang}.
The quantum Hall/superconductor/quantum Hall junctions can host the perfect CAR in the quantum limit with the zeroth Landau levels being involved\cite{SBZhang}. A robust signature of the perfect CAR is also found in a superconductor sandwiched between two antiferromagnetic layers with one being electron doped and the other being hole doped\cite{Jakobsen}.

Except for the above researches, the graphene/superconductor/graphene junctions are another system to achieve CAR\cite{Benjamin2,Crepin,Beconcini}. Especially, LAR and ECT will be both eliminated in the absence of valley-isospin or spin polarization when the left and right graphene leads are of the $n$ type and the $p$ type, respectively\cite{Cayssol}. In the superconducting graphene spin valves, the switch between the perfect ECT and the perfect CAR can be created by reversing the magnetization direction in one of the ferromagnetic layers\cite{Linder1}. The mechanism for the perfect CAR in Ref.[\onlinecite{Linder1}] is to reduce the Fermi surface for minority spins to a single point through tuning the local Fermi level to be equivalent to the exchange splitting. When graphene is in the quantum Hall regime under a high magnetic field, an exclusive CAR without other Andreev reflections can also be obtained based on the chiral edge states\cite{Hou,expGS}.

In this paper, we propose a distinct mechanism to achieve the spin-valley dependent perfect CAR in the proximitized graphene/superconductor/proximitized graphene junctions.
In our junctions, graphene possesses the pseudospin staggered potential and the intrinsic spin-orbit coupling induced by the proximity effect of substrates\cite{Zihlmann,Zollner1,Khatibi,Zollner2,Wakamura,Frank1}.
When the intrinsic spin-orbit couplings in the left graphene and the right graphene have the opposite sign, the junctions are insulating for the electrons with specific types of the spin-valley indices.
The perfect CAR with LAR and ECT being completely eliminated can happen for the electrons with the other types of the spin-valley indices. Furthermore, the CAR can occur in the whole superconducting gap under the given pseudospin staggered potential and the intrinsic spin-orbit coupling. The CAR probability can reach $100\%$ in the space composed of the incident angle and the energy of the electrons. The crossed conductance spectra and the oscillation of the crossed conductance with the superconductor length are investigated. The energy ranges for the CAR without LAR and ECT under different magnitudes of the potential and the coupling are clarified. If one changes the sign of the intrinsic spin-orbit coupling or exchanges the biases applied on the left graphene and the right graphene, the spin-valley indices of the electrons for the insulating state and the perfect CAR will be exchanged. Our proposal for the CAR without LAR and ECT provides a feasible approach for the highly efficient Cooper pair splitting.

The rest of the paper is organized as follows. In Sec.\uppercase\expandafter{\romannumeral 2}, we give the model of the proximitized graphene/superconductor/proximitized graphene junctions and the formalism for the CAR probability and the crossed conductance. In Sec.\uppercase\expandafter{\romannumeral 3}, we present the numerical results and discussions about the CAR probability, the crossed conductance and the energy range for the CAR without LAR and ECT. Section.\uppercase\expandafter{\romannumeral 4} concludes this paper.

\section{\label{sec2}Model and Formulation}

\begin{figure}[!htb]
\centerline{\includegraphics[width=0.9\columnwidth]{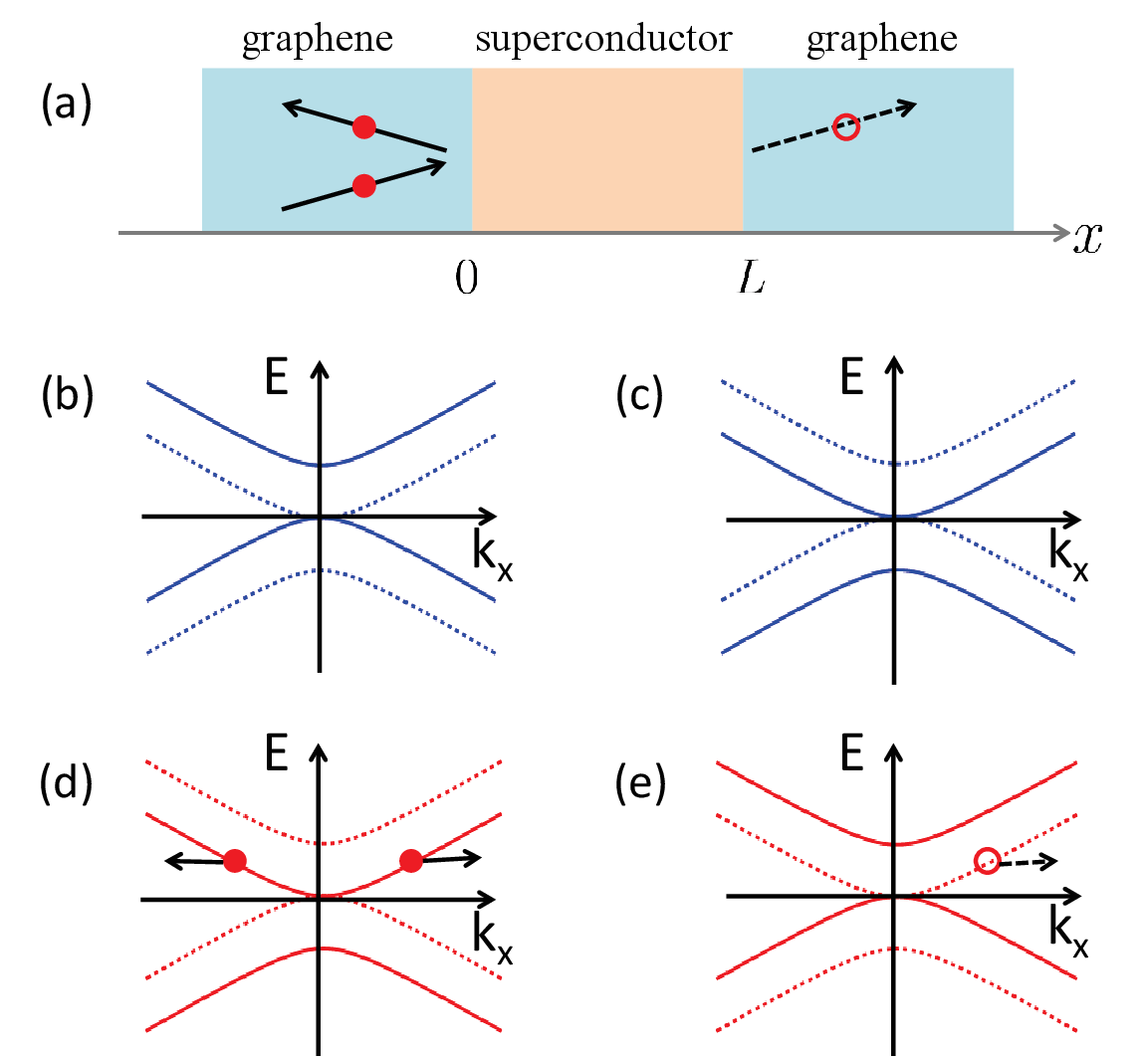}}
\caption{(a) The schematic illustration of the proximitized graphene/sperconductor/proximitized graphene junctions. The dispersions for the $K\uparrow$ electrons (blue solid lines) and the $K'\downarrow$ holes (blue dashed lines) in (b) the left graphene and (c) the right graphene. The dispersions for the $K\downarrow$ electrons (red solid lines) and the $K'\uparrow$ holes (red dashed lines) in (d) the left graphene and (e) the right graphene. The red solid (hollow) circle denotes the $K\downarrow$ electron ($K'\uparrow$ hole) and the black solid (dashed) arrows denote its motion direction. The scattering processes for the CAR without LAR and ECT of the $K\downarrow$ electrons are shown in (a),(d) and (e). The parameters $\Delta=\lambda>0$ have been taken for the dispersions.}
\label{fig1}
\end{figure}

The proximitized graphene/superconductor/proximitized graphene junctions
we consider in the $xy$ plane are schematically shown in Fig. {\ref{fig1}}(a).
The interfaces of the junctions are located at $x=0$ and $x=L$.
The electric transport is along the $x$ axis. The pseudospin staggered potential and the intrinsic spin-orbit coupling are induced in the left graphene and the right graphene by the substrates under them, which are denoted by $\Delta_{L},\lambda_{L}$ and $\Delta_{R},\lambda_{R}$, respectively.
The superconductivity in the middle region can be caused by a bulk superconductor through the proximity effect. The absences of the pseudospin potential and the intrinsic spin-orbit coupling are assumed in the superconducting region.

The Bogoliubov-de Gennes (BdG) Hamiltonian for the left (right) graphene can be written as\cite{Wang2,Cheng}
\begin{eqnarray}
H_{L(R)}=\tau_{z}\otimes h_{L(R)},\label{gBdG}
\end{eqnarray}
with
$h_{L(R)}=\hbar v_{F}(\xi k_{x}\sigma_{x}+k_{y}\sigma_{y})+\Delta_{L(R)}\sigma_{z}+\xi s_{z}\lambda_{L(R)}\sigma_{0}.$
Here the first term is for the pristine graphene which has the linear dispersions. The second and third terms are the pseudospin staggered potential $\Delta_{L(R)}$ and the spin-orbit coupling $\lambda _{L(R)}$, respectively.
${\bf{k}}=(k_{x},k_{y})$ is the momentum of particles in the $xy$ plane, $\xi=+(-)1$ for the $K(K')$ valley, $s_{z}=+(-)1$ for the up (down) spin, $\tau_{z}$ is the Pauli matrix in the particle-hole space, $\sigma_{x}$, $\sigma_{y}$, $\sigma_{z}$ and $\sigma_{0}$ are the Pauli matrices and the identity matrix in the pseudospin space.
The dispersions of the electrons with the specific spin-valley index can be solved from the BdG equation as
\begin{eqnarray}
E^{\pm}_{eL(R)\xi s_{z}}=\pm\sqrt{\hbar^2v_{F}^2(k_{x}^2+k_{y}^2)+\Delta_{L(R)}^2}+\xi s_{z}\lambda_{L(R)},\label{Eepm}
\end{eqnarray}
and those for the holes can be solved as
\begin{eqnarray}
E^{\pm}_{hL(R)\xi s_{z}}=\pm\sqrt{\hbar^2v_{F}^2(k_{x}^2+k_{y}^2)+\Delta_{L(R)}^2}-\xi s_{z}\lambda_{L(R)}.\label{Ehpm}
\end{eqnarray}
Here, $E^{+}_{eL(R)\xi s_{z}}$ and $E^{-}_{eL(R)\xi s_{z}}$ are respectively the conduction band and the valence band for electrons in the left (right) graphene while $E^{+}_{hL(R)\xi s_{z}}$ and $E^{-}_{hL(R)\xi s_{z}}$ are the valence band and the conduction band for holes in the left (right) graphene.
From Eqs. (\ref{Eepm}) and (\ref{Ehpm}),
one can find the linear dispersions at the zero staggered potential or the quadratic dispersions at the finite staggered potential.

The BdG Hamiltonian for the middle superconductor is given by\cite{Beenakker,Asano,Gao}
\begin{eqnarray}
H_{S}=\tau_{z}\otimes h_{S}+s_{z}\tau_{x}\otimes \tilde{\Delta}_{0},\label{HS}
\end{eqnarray}
with $h_{S}=\hbar v_{F}(\xi k_{x}\sigma_{x}+k_{y}\sigma_{y})-\mu_{s}\sigma_{0}$ and $\tilde{\Delta}_{0}=\Delta_{0}1_{2\times2}$. Here, $\mu_{s}$ is the chemical potential in superconductor and $\Delta_{0}$ is the magnitude of the superconducting gap. In Eq.(\ref{HS}), the pairing potential in superconductor couples the electron and the hole with the opposite spin-valley index. The eigenvalues for the quasiparticles in superconductor can be solved from the BdG equation as $E^{\pm}=\pm\sqrt{(\hbar v_{F}k-\mu_{s})^2+\Delta_{0}^2}$.

In this paper, we consider the symmetric case with
the equal left and right pseudospin staggered potential
$\Delta_{L}=\Delta_{R}=\Delta$ and
the opposite left and right intrinsic spin-orbit coupling
$\lambda_{L}=-\lambda_{R}=\lambda$.
For the given $\Delta$, the pseudohelical edge states, the quantum anomalous Hall effects and the chiral Majorana fermions in the proximitized graphene with $\lambda$ changing from the positive value to the negative value have been systematically studied in Refs.[{\onlinecite{Frank2,Hogl1,Hogl2}}].
Here we focus on the transport properties of
the bulk conduction and valence bands.
We take $\Delta_{L}=\Delta_{R}$ and $\lambda_{L}=-\lambda_{R}$ which will bring about the symmetric dispersions of electrons and holes.
The dispersions in the left graphene and the right graphene are schematically shown in Figs.{\ref{fig1}}(b)-(e) for $\Delta=\lambda>0$. The $K\uparrow$ electrons ($K'\downarrow$ holes) in the left graphene have the same dispersions with the $K\downarrow$ electrons ($K'\uparrow$ holes) in the right graphene. The $K\uparrow$ electrons ($K'\downarrow$ holes) in the left graphene also have the same dispersions with the $K'\downarrow$ ($K\uparrow$ electrons) holes in the right graphene and with the $K'\uparrow$ holes ($K\downarrow$ electrons) in the left graphene.

For $\Delta=\lambda$, the bands involved in the electric transport along the $+x$ axis are the conduction bands for electrons and the valence bands for holes. Since the time-reversal symmetry is still satisfied in our system, the $K\uparrow(K\downarrow)$ electrons and $K'\downarrow(K'\uparrow)$ electrons are degenerate. We only consider the scattering processes for the injection of the $K\uparrow$ and the $K\downarrow$ electrons.
For the injection of a $K\uparrow$ electron, the wave functions in the three regions can be solved from the BdG equations as
\begin{align}
\label{kuL}\Psi_{L\uparrow}(x<0)=\psi^{+}_{Le\uparrow}e^{ik_{x}^{Le\uparrow}x}+r_{e1}\psi^{-}_{Le\uparrow}e^{-ik_{x}^{Le\uparrow}x}\nonumber\\
+r_{h1}\psi^{-}_{Lh\downarrow}e^{-ik_{x}^{Lh\downarrow}x},\\
\label{kuS}\Psi_{S\uparrow}(0<x<L)=f_{1}\psi^{+}_{e\uparrow}e^{ik^{e\uparrow}_{x}x}+f_{2}\psi^{-}_{e\uparrow}e^{-ik^{e\uparrow}_{x}x}\nonumber\\
+f_{3}\psi^{+}_{h\downarrow}e^{ik_{x}^{h\downarrow}x}+f_{4}\psi_{h\downarrow}^{-}e^{-ik_{x}^{h\downarrow}x},\\
\label{kuR}\Psi_{R\uparrow}(x>L)=t_{e1}\psi^{+}_{Re\uparrow}e^{ik^{Re\uparrow}_{x}x}+t_{h1}\psi^{+}_{Rh\downarrow}e^{ik^{Rh\downarrow}_{x}x},
\end{align}
with $\psi^{+}_{L(R)e\uparrow}=(\chi_{L(R)1}\eta_{L(R)1},1,0,0)^{T}$, $\psi^{-}_{Le\uparrow}=(-\chi_{L1}\eta^{*}_{L1},1,0,0)^{T}$, $\psi^{-(+)}_{L(R)h\downarrow}=(0,0,\chi_{L(R)2}\eta_{L(R)2},1)^{T}$, $\psi_{e\uparrow}^{+}=(u\eta_{s1},u,v\eta_{s1},v)^{T}$, $\psi_{e\uparrow}^{-}=(-u\eta_{s1}^{*},u,-v\eta_{s1}^{*},v)^T$, $\psi_{h\downarrow}^{+}=(v\eta_{s2},v,u\eta_{s2},u)^T$ and $\psi_{h\downarrow}^{-}=(-v\eta_{s2}^{*},v,-u\eta_{s2}^{*},u)^T$. Here,
$\chi_{L(R)1}=[\Delta-(+)\lambda+E]/\sqrt{[E-(+)\lambda]^2-\Delta^2}$,
$\chi_{L(R)2}=[\Delta-(+)\lambda-E]/\sqrt{[E+(-)\lambda]^2-\Delta^2}$,
$\eta_{L(R)1}=[k_{x}^{L(R)e\uparrow}-i k_{y}]/{k_{L(R)e\uparrow}}$,
$\eta_{L(R)2}=[-(+)k_{x}^{L(R)h\downarrow}-i k_{y}]/{k_{L(R)h\downarrow}}$,
$\eta_{s1}=(k_{x}^{e\uparrow}-i k_{y})/{k_{e\uparrow}}$ and
$\eta_{s2}=(k_{x}^{h\downarrow}-i k_{y})/{k_{h\downarrow}}$.
The wave vectors are $k_{y}=\sqrt{(E-\lambda)^2-\Delta^2}\sin{\theta}/\hbar v_{F}$,
$k_{L(R)e\uparrow}=\sqrt{[E-(+)\lambda]^2-\Delta^2}/\hbar v_{F}$,
$k_{L(R)h\downarrow}=\sqrt{[E+(-)\lambda]^2-\Delta^2}/\hbar v_{F}$,
$k_{x}^{L(R)e\uparrow}=\sqrt{k_{L(R)e\uparrow}^2-k_{y}^2}$ and
$k_{x}^{L(R)h\downarrow}=\sqrt{k_{L(R)h\downarrow}^2-k_{y}^2}$.
Here, $\theta$ is the incident angle for the $K\uparrow$ electrons from the left graphene. In superconductor, we consider the heavily doped case with $\mu_{s}>>(\Delta_{0},\Delta,\lambda)$, the wave vectors can be expressed as $k_{x}^{e\uparrow(h\downarrow)}=[\mu_{s}+(-)\sqrt{E^2-\Delta_{0}^2}/2]/\hbar v_{F}$ for $E>\Delta_{0}$ and $k_{x}^{e\uparrow(h\downarrow)}=[\mu_{s}+(-)i\sqrt{\Delta_{0}^2-E^2}/2]/{\hbar v_{F}}$ for $E<\Delta_{0}$. The coherent factors $u=\sqrt{(E+\sqrt{E^2-\Delta_{0}^2})/{2E}}$ and $v=\sqrt{(E-\sqrt{E^2-\Delta_{0}^2)}/{2E}}$.

For the injection of a $K\downarrow$ electron, the wave functions are given by
\begin{align}
\label{kdL}\Psi_{L\downarrow}(x<0)=\psi^{+}_{Le\downarrow}e^{ik_{x}^{Le\downarrow}x}
+r_{e2}\psi^{-}_{Le\downarrow}e^{-ik_{x}^{Le\downarrow}x}\\\nonumber
+r_{h2}\psi^{-}_{Lh\uparrow}e^{-ik_{x}^{Lh\uparrow}x},\\
\label{kds}\Psi_{S\downarrow}(0<x<L)=f_{5}\psi^{+}_{e\downarrow}e^{ik^{e\downarrow}_{x}x}
+f_{6}\psi^{-}_{e\downarrow}e^{-ik^{e\downarrow}_{x}x}\\\nonumber
+f_{7}\psi^{+}_{h\uparrow}e^{ik_{x}^{h\uparrow}x}+f_{8}\psi_{h\uparrow}^{-}e^{-ik_{x}^{h\uparrow}x},\\
\label{kdR}\Psi_{R\downarrow}(x>L)=t_{e2}\psi^{+}_{Re\downarrow}e^{ik^{Re\downarrow}_{x}x}
+t_{h2}\psi^{+}_{Rh\uparrow}e^{ik^{Rh\uparrow}_{x}x}.
\end{align}
The expression of $\Psi_{L(R)\downarrow}$ can be obtained by substituting $-\lambda$ for $\lambda$ in $\Psi_{L(R)\uparrow}$. The expression of $\Psi_{S\downarrow}$ can be obtained by substituting $-u$ for $u$ in $\psi^{\pm}_{e\uparrow}$ and substituting $-v$ for $v$ in $\psi^{\pm}_{h\uparrow}$. More details for the solving processes of the wave functions in Eqs.(\ref{kuL})-(\ref{kdR}) can be found in Appendix.

Using the following boundary conditions,
\begin{eqnarray}
\Psi_{L\uparrow(\downarrow)}(x=0^{-})=\Psi_{S\uparrow(\downarrow)}(x=0^{+}),\\
\Psi_{S\uparrow(\downarrow)}(x=L^{-})=\Psi_{R\uparrow(\downarrow)}(x=L^{+}),
\end{eqnarray}
we can solve the normal reflection coefficients $r_{e1}$ and $r_{e2}$, the LAR coefficients $r_{h1}$ and $r_{h2}$, the ECT coefficients $t_{e1}$ and $t_{e2}$ and the CAR reflection coefficients $t_{h1}$ and $t_{h2}$. The corresponding probabilities can be expressed as
\begin{align}
R_{e1(2)}&=\vert r_{e1(2)}\vert^2,\\
R_{h1(2)}&=-\frac{k_{x}^{Lh\downarrow(\uparrow)}k_{Le\uparrow(\downarrow)}\chi_{L2}(+(-)\lambda)}
{k_{x}^{Le\uparrow(\downarrow)}k_{Lh\downarrow(\uparrow)}\chi_{L1}(+(-)\lambda)}\vert r_{h1(2)}\vert^2,\\
T_{e1(2)}&=\frac{k_{x}^{Re\uparrow(\downarrow)}k_{Le\uparrow(\downarrow)}\chi_{R1}(+(-)\lambda)}
{k_{x}^{Le\uparrow(\downarrow)}k_{Re\uparrow(\downarrow)}\chi_{L1}(+(-)\lambda)}\vert t_{e1(2)}\vert^2,\\
T_{h1(2)}&=-\frac{k_{x}^{Rh\downarrow(\uparrow)}k_{Le\uparrow(\downarrow)}\chi_{R2}(+(-)\lambda)}
{k_{x}^{Le\uparrow(\downarrow)}k_{Rh\downarrow(\uparrow)}\chi_{L1}(+(-)\lambda)}\vert t_{h1(2)}\vert^2.
\end{align}
They satisfy the condition of the probability conservation, i.e.,$R_{e1(2)}+R_{h1(2)}+T_{e1(2)}+T_{h1(2)}=1$. According to the Blonder-Tinkham-Klapwijk theory\cite{BTK}, the normalized crossed conductance at the zero temperature for the bias voltage $V$ is given
by\cite{Linder1,Linder2}
\begin{eqnarray}
G_{CAR}(eV)=\frac{G_{\uparrow CAR}+G_{\downarrow CAR}}{G_{\uparrow0}+G_{\downarrow0}},\label{Gcar}
\end{eqnarray}
where $G_{\uparrow(\downarrow) CAR}=\frac{e^2}{h}N_{\uparrow(\downarrow)}(eV)\int_{-\pi/2}^{\pi/2}T_{h1(2)}\cos{\theta}d\theta$ and $G_{\uparrow(\downarrow)0}=\frac{e^2}{h}N_{\uparrow(\downarrow)}(eV)$ with the number of the transverse modes $N_{\uparrow(\downarrow)}(E)=W\sqrt{[E-(+)\lambda]^2-\Delta^2}/\pi\hbar v_{F}$ in the left graphene. Here, $W$ is the width of the left graphene, which is also the width of the junctions.
When the ECT process disappears, the crossed conductance is equal to the nonlocal conductance
$dI_L/dV_R$ with $I_L$ the current in the left graphene and $V_R$ the bias of the right graphene terminal\cite{SBZhang,nref1,nref2}.

\section{\label{sec3}Results and discussions}
In our calculations, we take the superconducting gap $\Delta_{0}$ as the unit of the energy $E$, the chemical potential $\mu_{s}$, the staggered potential $\Delta$ and the spin-orbit coupling strength $\lambda$. The unit of the length $L$ of superconductor is taken as $\xi_{0}=\hbar v_{F}/\Delta_{0}$. For the chemical potential $\mu_{s}$ in superconductor, we chose $\mu_{s}=100\Delta_{0}$ which ensures the quasiparticles in superconductor propagating parallel to the $x$ axis as described by the wave functions in Eqs.(\ref{kuS}) and (\ref{kds}). In addition, we will only present and discuss the numerical results for the injection of the $K\uparrow$ and the $K\downarrow$ electrons for simplicity since the they are degenerate with the $K'\downarrow$ and the $K'\uparrow$ electrons due to the time-reversal symmetry. The latter two types of electrons have the same results with the former ones.

We first consider the situation of $\Delta=\lambda>0$.
From Eq.(\ref{Eepm}), one can find there is a gap
from $E=0$ to $E=2\Delta$ for the $K\uparrow$ electrons in the left graphene
[see Fig.{\ref{fig1}}(b)].
The junctions are insulating for the $K\uparrow$ electrons in the energy range $0<E<2\Delta$ when the electric transport along the $+x$ direction is considered.
However, for the $K\downarrow$ electrons, the junctions are conductive
in the energy range $0<E<2\Delta$.
Taking the same gaps for the $K'\uparrow$ holes in the left graphene
and the $K\downarrow$ electrons in the right graphene into account,
LAR and ECT will not happen when the $K\downarrow$ electrons are injected
from the left graphene in the energy range $0<E<2\Delta$.
In other words, the CAR without LAR and ECT can be realized for the $K\downarrow$ electrons in this energy range for our junctions.
The scattering process in the energy range $0<E<2\Delta$ with only CAR and
normal reflection for the $K\downarrow$ electrons is schematically shown in Fig.{\ref{fig1}}(a) or Figs.{\ref{fig1}}(d) and 1(e).
When $E>2\Delta$, both the $K\uparrow$ and the $K\downarrow$ electrons
will anticipate the transport. The LAR and ECT for the electrons will be activated.

\begin{figure}[htbp]
\centering
\includegraphics[width=1\columnwidth]{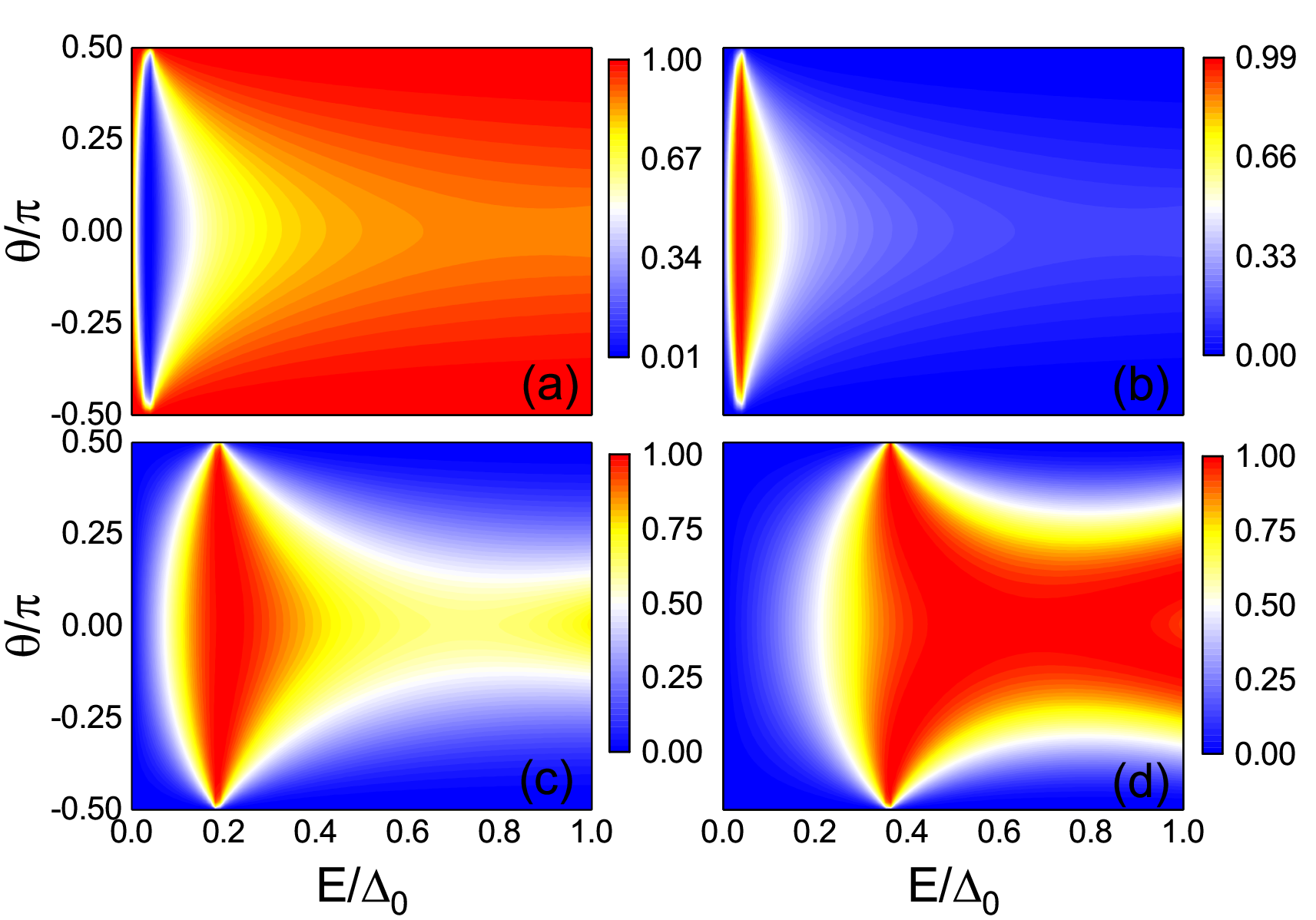}
\caption{(a) The normal reflection probability and (b) the CAR probability of the incident $K\downarrow$ electrons for $L=0.55\xi_{0}$ in the $(E,\theta)$ space.
The CAR probabilities of the incident $K\downarrow$ electrons for (c) $L=1.43\xi_{0}$
and (d) $L=2.53\xi_{0}$ in the $(E,\theta)$ space. The parameters are taken as $\Delta_L=\Delta_R=0.5\Delta_{0}$ and $\lambda_{L}=-\lambda_{R}=0.5\Delta_{0}$.}\label{fig2}
\end{figure}

\begin{figure}[htbp]
\centering
\subfigure{
\begin{minipage}{0.9\columnwidth}
\centering
\includegraphics[width=0.9\columnwidth]{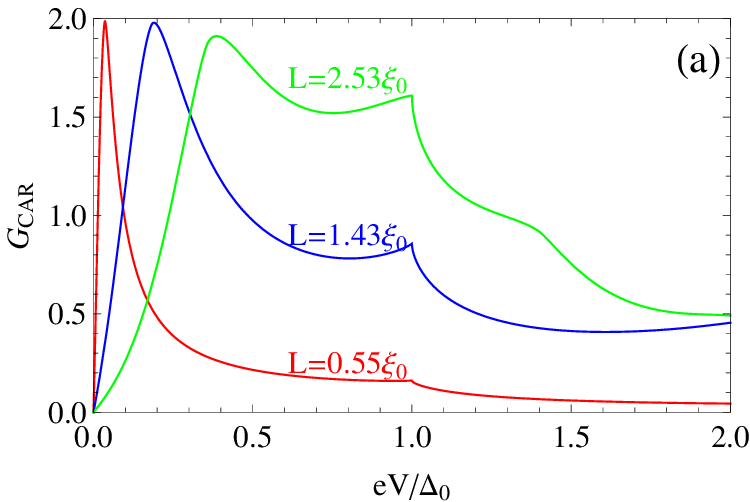}
\end{minipage}
}
\subfigure{
\begin{minipage}{0.9\columnwidth}
\centering
\includegraphics[width=0.9\columnwidth]{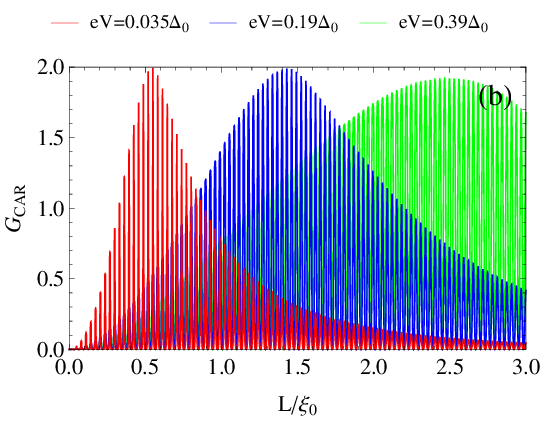}
\end{minipage}
}
\caption{(a) The crossed conductance spectra and (b) the oscillations of the crossed conductances with the length of superconductor for the injection of the $K\downarrow$ electrons. Other parameters are taken as the same with those in Fig.{\ref{fig2}}.}\label{fig3}
\end{figure}

In order to present the numerical results, we take the values of $\Delta$ and $\lambda$ as $\Delta=\lambda=0.5\Delta_{0}$.
In this situation,
there will be an energy gap spanning from $E=0$ to $E=2\Delta=\Delta_{0}$
for the $K'\uparrow$ holes in the left graphene
and the $K\downarrow$ electrons in the right graphene, as shown
in Figs. {\ref{fig1}}(d) and {\ref{fig1}}(e).
In other words, the local holes responsible for LAR and the nonlocal electrons responsible for ECT for the incidence of the $K\downarrow$ electrons from the left graphene are absent.
So the CAR without LAR and ECT for the injection of the $K\downarrow$ electrons can happen in the whole superconducting gap, i.e., the energy range $0<E<\Delta_{0}$.
Since our junctions are insulating for the $K\uparrow$ electrons in the same energy range, we will only present the numerical results of the reflection probabilities for the $K\downarrow$ electrons.
In Fig.{\ref{fig2}}, we show the normal reflection probability $R_{e2}$ and the CAR probability $T_{h2}$ in the $(E,\theta)$ space for different superconductor lengthes.
Here $E$ is the energy of the incident electron,
which can experimentally be controlled by the gate voltage
or the bias\cite{nature11,nature12}.
Three situations of the superconductor length are considered, which are $L<\xi_{0} (L=0.55\xi_{0})$, $L\sim\xi_{0} (L=1.43\xi_{0})$ and $L>\xi_{0} (L=2.53\xi_{0})$.
For $L=0.55\xi_{0}$, the normal reflection probability $R_{e2}$ is nearly equal to zero around $E=0.035\Delta_{0}$ as shown in Fig.\ref{fig2}(a). Conversely, the CAR process mainly happens around $E=0.035\Delta_{0}$ and is almost independent on the incident angle $\theta$ as shown in Fig.\ref{fig2}(b).
The maximum value of the CAR probability $T_{h2}$ can reach $99\%$.
Because the LAR and ECT processes are absent in the superconducting gap,
the conservation relation $R_{e2}+T_{h2}=1$ is satisfied for $0<E<\Delta_{0}$.
For $L=1.43\xi_{0}$ and $L=2.53\xi_{0}$, we only present the CAR probabilities in Figs.\ref{fig2} (c) and (d), respectively, and do not give the normal reflection probabilities due to the presence of the probability conservation.
We find that when $L$ is increased, the dominant zone for the CAR process in the $(E,\theta)$ space will be enlarged and will stay further away from $E=0$.
For $L=1.43\xi_{0}$, the nearly $\theta$-independent CAR happens around $E=0.2\Delta_{0}$ while for $L=2.53\xi_{0}$, it happens around $E=0.4\Delta_{0}$. Furthermore, the maximum value of the CAR probability $T_{h2}$ can reach $100\%$ for the both cases.
The perfect CAR along with the absences of the normal reflection, LAR and ECT processes will be realized in the $(E,\theta)$ space.
It also is worth mentioning that the CAR probability oscillates rapidly
with the change of the length $L$ [e.g. see Fig.\ref{fig3}(b)].
For some length values, the CAR probability $T_{h2}$ does not reach $100\%$.
Here, we emphasize that even if $T_{h2}$ is much smaller than $1$,
the CAR process without LAR and ECT happens still.
So when a Cooper pair is split, two electrons must respectively go to
the left and right graphene leads still.

In Fig.\ref{fig3}(a), we give the normalized crossed conductance spectra for the injection of the $K\downarrow$ electrons through integrating $T_{h2}$ about the incident angle according to Eq.(\ref{Gcar}).
In Eq.(\ref{Gcar}), $G_{\uparrow CAR}$ and $G_{\uparrow 0}$ are equal to zero for $0<E<\Delta_{0}$ since the transport of the $K\uparrow$ electrons is absent.
However, $G_{\uparrow CAR}$ and $G_{\uparrow 0}$ are finite and contribute to the normalized crossed conductance for $E>\Delta_{0}$.
For $L=0.55\xi_{0}$, the maximum value almost being $2$ of the crossed conductance $G_{CAR}$ can be obtained at about $E=0.035\Delta_{0}$.
This is because the maximum value about $99\%$ of the CAR probability distributes
nearly along the vertical line with $E=0.035\Delta_{0}$ in the $(E,\theta)$ space as shown in Fig.\ref{fig2}(b).
For $L=1.43\xi_{0}$, the maximum value of the crossed conductance can still reach $G_{CAR}\approx2$ at about $E=0.19\Delta_{0}$.
It is weakened a little compared with that for $L=0.55\xi_{0}$
although the CAR probability for $L=1.43\xi_{0}$ can amount to $100\%$.
This is because the dominate zone for CAR will be enlarged and extend towards right when $L$ is increased as discussed above.
The $100\%$ probability of CAR for $L=1.43\xi_{0}$ will not gather along a vertical line with the definite energy in the $(E,\theta)$ space as shown in Fig.\ref{fig2}(c).
This point can be seen more clearly from the CAR probability in Fig.\ref{fig2}(d)
and the crossed conductance in Fig.\ref{fig3}(a) for $L=2.53\xi_{0}$.
In this situation, the dominate zone for CAR can extend from about $E=0.3\Delta_{0}$ to $E=\Delta_{0}$. Accordingly, the crossed conductance $G_{CAR}$ is weakened more obviously. Even so, the crossed conductance for $L=2.53\xi_{0}$ can still attain its value exceeding $1.5$ in a wide energy range. In addition, the crossed conductances in Fig.\ref{fig3}(a) are always equal to zero for $eV=0$ because the number of the transverse modes $N_{\downarrow}$ in Eq.(\ref{Gcar}) for the $K\downarrow$ electrons will vanish at $E=0$ for $\Delta=\lambda$.

In Fig.\ref{fig3}(b), the variations of $G_{CAR}$ with the superconductor length $L$ are presented. The bias voltages are taken as $eV=0.035\Delta_{0}$, $0.19\Delta_{0}$ and $0.39\Delta_{0}$, which correspond to the positions of the peak values for $G_{CAR}$ in Fig.\ref{fig3}(a). From Fig.\ref{fig3}(b), one can find the crossed conductances exhibit the intense oscillations with the superconductor length. The envelop lines of $G_{CAR}$ first increase from zero when the superconductor length is raised from $L=0$.
After reaching their peak values, the lines will decrease as the length continues to rise. For the large enough length, the crossed conductances will become zero again.
On the other hand, the position for the peak value of $G_{CAR}$ in Fig.\ref{fig3}(b) will move right when the bias voltage is increased and the peak value will be weakened. This is consistent with the result for the crossed conductance spectra in Fig.\ref{fig3}(a).

\begin{figure}[htbp]
\centering
\subfigure{
\begin{minipage}{0.9\columnwidth}
\centering
\includegraphics[width=0.9\columnwidth]{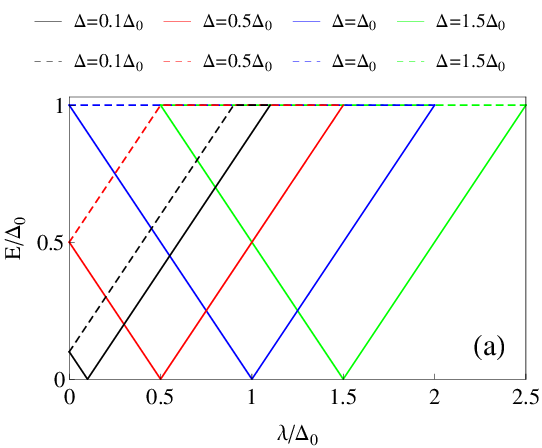}
\end{minipage}
}
\subfigure{
\begin{minipage}{0.9\columnwidth}
\centering
\includegraphics[width=0.9\columnwidth]{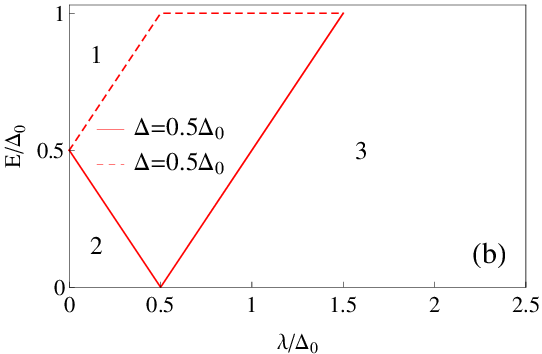}
\end{minipage}
}
\caption{(a) The energy ranges for the CAR without LAR and ECT for the injection of the $K\downarrow$ electrons. The solid lines are the lower bounds while the dashed lines are the upper bounds for the different values of $\Delta$.
(b) The three zones surrounding the energy range for the CAR without LAR and ECT with $\Delta=0.5\Delta_{0}$.}\label{fig4}
\end{figure}

The occurrence of the CAR without LAR and ECT for the $K\downarrow$ electrons in our junctions is not limit to the case of $\Delta=\lambda$.
We now turn to the situation of $\Delta\ne\lambda$.
In this case, the symmetries of the energy bands presented in Fig.{\ref{fig1}} still hold. Here $\Delta>0$ and $\lambda>0$ are still assumed.
If $\Delta-\lambda>\Delta_{0}$, there will be a gap from $E=-\Delta-\lambda<0$ to $E=\Delta-\lambda>\Delta_{0}$ for the $K\downarrow$ electrons. The junctions will be insulating for the $K\downarrow$ electrons in the energy range $0<E<\Delta_{0}$.
The CAR can not be expected in the superconducting gap in our junctions.
However, for $\Delta-\lambda<\Delta_{0}$, the CAR without LAR and ECT for the $K\downarrow$ electrons can happen in the energy range $E_{min}<E<E_{max}$ inside the superconducting gap. We can derive $E_{min}=\vert\Delta-\lambda\vert$ and $E_{max}=\text{min}(\Delta+\lambda,\Delta_{0})$ from Eqs.(\ref{Eepm}) and (\ref{Ehpm}).
Fig.\ref{fig4}(a) shows the energy ranges for the CAR without LAR and ECT for the different values of the pseudospin potential $\Delta$.
For each $\Delta$, the lower bound (the solid line) of the energy range is determined by $E_{min}$ while the upper bound (the dashed line) is determined by $E_{max}$.
As long as $\Delta\ge0.5\Delta_{0}$, the CAR without LAR and ECT in the whole superconducting gap can be realized at $\lambda=\Delta$, which can be seen from the zones surround by the red, blue and green lines for $\Delta=0.5\Delta_{0}$, $\Delta_{0}$ and 1.5$\Delta_{0}$, respectively.
Otherwise the size of the energy range for the CAR will be smaller than the superconducting gap $\Delta_{0}$, i.e., $E_{max}-E_{min}<\Delta_{0}$ for all values of $\lambda$, which can be seen from the zone surround by the black lines for $\Delta=0.1\Delta_{0}$. In addition, if $\Delta<\Delta_{0}$, the zone for the CAR without LAR and ECT shows itself as a trapezoid. As $\Delta$ is increased, the area of the trapezoid becomes larger. When $\Delta$ increases up to $\Delta_{0}$, the zone will become an inverted triangle. The larger value of $\Delta$ no longer changes the triangle shape of the zone but will move the triangle zone to the right.
The inverted triangle zone for the CAR without LAR and ECT is symmetric about $\lambda=\Delta$. These characters of the energy ranges for the occurrence of the CAR without LAR and ECT can be seen obviously from Fig.\ref{fig4}(a). In these energy ranges in Fig.\ref{fig4}(a), the junctions are insulating for the $K\uparrow$ electrons.

In the other zones in the $(E,\lambda)$ space except for the trapezoid or the inverted triangle, the CAR without LAR and ECT for the $K\downarrow$ electrons will not happen. However, for different zones, the situations are very different.
We take $\Delta=0.5\Delta_{0}$ in Fig.\ref{fig4}(b) as an example.
There are three zones outside the trapezoid.
The zone $1$ corresponds to the energy range $\Delta+\lambda<E<\Delta_{0}$.
In this range, the CAR process still exists but the LAR and the ECT for the injection of the $K\downarrow$ electrons will be also activated. The incident $K\downarrow$ electrons belong to their conduction band in the left proximitized graphene, which tunnel to the conduction band of the electrons in the right proximitized graphene in the ECT process. The holes come from their valence band in the left (right) graphene in the LAR (CAR) process.
The zone $2$ corresponds to the energy range $0<E<\Delta-\lambda$, which is a gap for the $K\downarrow$ electrons. The junctions are insulating for the $K\downarrow$ electrons in the left proximitized graphene and CAR does not exist.
The zone $3$ corresponds to the energy range $0<E<-\Delta+\lambda$. In this range, the LAR and ECT for the $K\downarrow$ electrons will be activated.
The incident $K\downarrow$ electrons also belong to their conduction band, but they tunnel to the valence band of the electrons in the right proximitized graphene in the ECT process. The holes come from their conduction (valence) band in the left (right) proximitized graphene in the LAR (CAR) process.
Here, we do not analyze the scattering processes for the $K\uparrow$ electron in the three energy zones since they will not provide the CAR without LAR and ECT either. The above analyses also apply to the cases for the other values of the pseudospin potential $\Delta$ in Fig.\ref{fig4}(a).

In the above results and discussions, we have assumed $\Delta>0$ and $\lambda>0$.
For $\Delta>0$ and $\lambda<0$, the energy band structures for the $K\downarrow$ electrons in Figs.\ref{fig1}(d) and (e) will exchange with those for the $K\uparrow$ electrons in Figs.\ref{fig1}(b) and (c).
The CAR without LAR and ECT will happen for the incident $K\uparrow$ electrons from the left proximitized graphene. The above results and discussions are also applicable.
It's just that the spin-valley index of the electrons for the CAR without LAR and ECT becomes $K\uparrow$. In other words, one can switch the spin-valley index of the electrons for the CAR by changing the sign of the intrinsic spin-orbit coupling $\lambda$.
Another alternative method to realize the conversion of the spin-valley index is to exchange the biases applied on the left graphene and the right graphene.
Due to the symmetric relations between the particle dispersions for the left graphene and those for the right graphene as shown in Figs.{\ref{fig1}}(b)-(e), the CAR without LAR and ECT for the incident $K\uparrow$ electrons from the right graphene can be realized when the electric transport is along the $-x$ axis. At this time, the dispersion for the $K\downarrow$ electrons in the right graphene opens a gap as shown in Fig.\ref{fig1}(e). The junctions are insulating for the $K\downarrow$ electrons in the transport along the $-x$ axis. In a word, our junctions possess the flexibility for the realization of the perfect CAR for the electrons with the specific spin-valley index.

Note that our junctions are different from the structure based on silicene in Ref.[\onlinecite{Linder2}]. There, the effective spin-orbit coupling is used to open gaps. The intrinsic spin-orbit coupling here in our junctions plays the role of the effective chemical potential which is dependent on the spin-valley index as shown in Eqs.(\ref{Eepm}) and (\ref{Ehpm}). In Ref.[\onlinecite{Linder2}], except for the staggered sublattice potential and the effective spin-orbit coupling, the gate voltages are also indispensable for the realization of CAR, which are used to adjust the spin-valley independent chemical potentials. There, the spin-valley index of the electrons for CAR can not be simply switched by changing the sign of the effective spin-orbit coupling or exchanging the biases applied on the left silicene and the right silicene. In addition, our junctions are distinct from
the $p$-type semiconductor/superconductor/$n$-type semiconductor
structure\cite{Veldhorst} and the quantum spin Hall insulator/superconductor/quantum spin Hall insulator structure\cite{Chen84}.
In these structures, there is no the valley index for particles,
which are different from the spin-valley dependent CARs in our junctions.

As the inverse process of CAR, the Cooper pair splitting can produce
two spatially separate electrons.
It is worth emphasizing that the state of the two splitting electrons
in our proximitized graphene/superconductor/proximitized graphene junction
is the spin-entangled and valley-entangled state.
For the proximitized graphene, the staggered potential
and the spin-orbit coupling do not break the time-reversal symmetry.
The $K\uparrow (K\downarrow)$ electrons and the $K'\downarrow (K'\uparrow)$
electrons are degenerate.
Both the incident $K\downarrow$ and $K'\uparrow$ electrons
from the left terminal are responsible for the CAR without LAR and ECT
under the parameters in Figs.\ref{fig2} and \ref{fig3}.
In other words, the involved electrons in the CAR possess
two different spin indices ($\uparrow$ or $\downarrow$)
and two different valley indices ($K$ or $K'$).
When the Cooper pair is split,
the $K$-valley spin-down electron can randomly go to both the left and right graphene terminals, and so can the $K'$-valley spin-up electron.
So the state of the two splitting electrons is
$\frac{\sqrt{2}}{2}\left(|LK\downarrow\rangle |RK'\uparrow\rangle +
|LK'\uparrow\rangle |RK\downarrow\rangle\right)$,
which is the spin-entangled and valley-entangled state.
So our junctions are distinct from the half metal/superconductor/half metal junctions which can not produce the entangled two-electron state.

Now, we give the discussions about the experimental realization of our junctions.
The condition for the achievement of the CAR without LAR and ECT
in the whole superconducting gap in our junctions is the opposite sign of the intrinsic spin-orbit couplings in the left graphene and the right graphene, i.e., $\lambda_{L}=-\lambda_{R}$.
Recent studies manifest the intrinsic spin-orbit coupling with the positive sign can be obtained in the graphene on WS$_{2}$\cite{Wang2} or in the graphene on the topological insulator Bi$_{2}$Se$_{3}$\cite{Zollner4}.
On the other hand, the intrinsic spin-orbit coupling with the negative sign can be obtained in the graphene on transition-metal dichalcogenides\cite{Gmitra1,Gmitra2}.
In Refs.[\onlinecite{Wang2}], the intrinsic spin-orbit coupling of the positive value $2.5meV$ is estimated. In Refs.[\onlinecite{Zollner4,Gmitra1,Gmitra2}], the intrinsic spin-orbit couplings are considered as the sublattice dependent quantities and are denoted by $\lambda^{A}_{I}$ and $\lambda^{B}_{I}$, but the relation $\vert\lambda^{A}_{I}\vert\approx\vert\lambda^{B}_{I}\vert$ always holds true.
In Ref.[\onlinecite{Zollner4}], $\lambda_{I}^{A}>0$ and $\lambda_{I}^{B}<0$ are predicted while in Refs.[\onlinecite{Gmitra1,Gmitra2}], $\lambda_{I}^{A}<0$ and $\lambda_{I}^{B}>0$ are claimed.
Since we have taken $\lambda=\lambda_{I}^{A}$ and $\lambda=-\lambda_{I}^{B}$ in the Hamiltonian for the proximitized graphene as given in Eq.(\ref{gBdG}), this indicates that $\lambda>0$ in Ref.[\onlinecite{Zollner4}] and $\lambda<0$ in Refs.[\onlinecite{Gmitra1,Gmitra2}] can be obtained. Especially, the existing experimental results such as those in Ref.[\onlinecite{Zihlmann,Wang2,ZWang}] can be well explained by the theoretical model for the proximitized graphene\cite{Zollner4,Gmitra1,Gmitra2}, which is just the one adopted by us to describe our junctions. Actually, the tunable spin-orbit coupling in graphene on WS$_2$ by adjusting the interlayer distance has been reported experimentally\cite{Yang3}. These existing studies about the proximitized graphene suggest that our junctions model is realistic in experiment.

In addition, it is worth mentioning that
the opposite sign and same magnitude of the spin-orbit coupling
in the left and right graphene leads are not the necessary condition
for the CAR without LAR and ECT in our junctions.
For example, if one takes $\Delta_L=\Delta_R=\Delta_0$ and $\lambda_L=2\lambda_R=0.5\Delta_0$, the CAR without LAR an ECT can be obtained in the energy range $0.75\Delta_0<E<\Delta_0$ for the injection of the $K\downarrow$ electrons.
Actually, the same magnitude of the staggered potentials in the left and right graphene leads is not necessary either.
The absence of the strong restrictions on the values of $\Delta_L$, $\Delta_R$, $\lambda_L$ and $\lambda_R$ can enhance the experimental flexibility and feasibility.

The magnitude of the superconducting gap will exert an upper limit of the energy range for observing the CAR without LAR and ECT in our junctions as shown in Fig.{\ref{fig4}}(a).
As for the magnitude of the superconducting gap induced in graphene, its value can be estimated as $0.1$ meV.
For example, the superconducting gap of $0.05$-$0.06$meV in graphene can be realized in the NbSe$_2$/graphene van der Waals junction\cite{Moriya}.
An alternative method to enhance the gap and to enlarge the energy range for CAR is to substitute a bulk superconductor for the middle region.
In fact, the direct contact between graphene and a bulk superconductor is a mature experimental technology\cite{Sahu}.
In this situation, the CAR without LAR and ECT can still be expected since the energy band structures in the left graphene and the right graphene are not changed.
\section{\label{sec4}Conclusions}

The CAR is studied in the proximitized graphene/superconductor/proximitized graphene junctions using the Blonder-Tinkham-Klapwijk theory.
The induced pseudospin staggered potential and the intrinsic spin-orbit coupling in the proximitized graphene are included in our junctions model. The CAR with LAR and ECT being simultaneously eliminated is achieved for the electrons with the specific spin-valley index. The perfect CAR with the probability of $100\%$ can be obtained in the space consisting of the incident angle and the energy of electrons. The crossed conductance and its oscillation with the superconductor length are investigated.
The energy ranges for the CAR without LAR and ECT are given for the different values of the potential and the coupling. The switch of the spin-valley indices of the electrons for the CAR can be realized easily, which provides the possibility for designing the highly efficient Cooper pair splitter based on the spin-valley degree of freedom.

\section*{\label{sec5}ACKNOWLEDGMENTS}
This work was financially supported
by the National Natural Science Foundation of China
(Grant No. 12374034, No. 11921005, and No. 11447175),
the Innovation Program for Quantum Science and Technology (2021ZD0302403),
the Strategic Priority Research Program of Chinese Academy of Sciences (XDB28000000)
and the project ZR2023MA005 supported by Shandong Provincial Natural Science Foundation.

\section*{APPENDIX}
\setcounter{equation}{0}
\renewcommand{\theequation}{A.\arabic{equation}}
Here, we present the derivations of the wave functions in Eqs.(\ref{kuL})-(\ref{kuR}) for the injection of the $K\uparrow$ electrons. For the wave function $\Psi_{L\uparrow}$ in the left graphene, it satisfies the following BdG equation
\begin{eqnarray}
H_{L}(-i\frac{\partial}{\partial x},k_y)\Psi_{L\uparrow}(x)=E\Psi_{L\uparrow}(x),
\end{eqnarray}
with the substitution of $k_x$ in $H_{L}$ in Eq.(\ref{gBdG})
with $-i\frac{\partial}{\partial x}$. The eigenvalue $E$ can be solved from the BdG equation, which have been given by Eqs.(\ref{Eepm}) and (\ref{Ehpm}). The $y$ component of the wave vector, i.e., $k_y$, is conserved in the scattering process due to the translation invariance of the junctions along the $y$ axis, which can be expressed as $k_y=k_{Le\uparrow}\sin{\theta}$. Here, the wave vector $k_{Le\uparrow}$ for the $K\uparrow$ electrons is solved from the eigenvalue $E^{+}_{eL+\uparrow}$ and $\theta$ is the incident angle of the $K\uparrow$ electrons. The wave vector $k_{Lh\downarrow}$ for the $K'\downarrow$ holes can be solve from $E^{+}_{hL-\downarrow}$. The solution of $\Psi_{L\uparrow}(x)$ can be expressed in Eq.(\ref{kuL}) with an incident wave of electrons, a reflected wave of electrons,
and a reflected wave of holes.
There is no approximation during the solving process of $\Psi_{L\uparrow}$. The wave function $\Psi_{R\uparrow}$ in the right graphene can be solved in a similar way.

For the wave function in the middle superconducting region, it satisfies the following BdG equation
\begin{eqnarray}
H_{S}(-i\frac{\partial}{\partial x},k_y)\Psi_{S\uparrow}(x)=E\Psi_{S\uparrow}(x).
\end{eqnarray}
The wave vectors $k_{e\uparrow}$ and $k_{h\downarrow}$ for the electron-like and the hole-like quasiparticles in superconductor can be solved from the eigenvalues $E^{\pm}=\pm\sqrt{(\hbar v_{F}k-\mu_s)^2+\Delta_{0}^{2}}$, which are given by $k_{e\uparrow(h\downarrow)}=\sqrt{\mu_s+(-)\sqrt{E^2-\Delta_0^2}}/\hbar v_{F}$. For the heavily doped case of $\mu_s>>(E,\lambda,\Delta)$, we have $k_{e\uparrow(h\downarrow)}>>k_{Le\uparrow}$. As a result, $k_{x}^{e\uparrow(h\downarrow)}=\sqrt{k_{e\uparrow(h\downarrow)}^2-k_y^2}\approx k_{e\uparrow(h\downarrow)}$ holds. Since we also have $\mu_{s}>>\sqrt{E^2-\Delta_0^2}$, $k_{x}^{e\uparrow(h\downarrow)}$ can be expanded as $[\mu_s+(-)\sqrt{E^2-\Delta_0^2}/2]/\hbar v_{F}$ for $E>\Delta_0$ and $[\mu_s+(-)i\sqrt{\Delta_0^2-E^2}/2]/\hbar v_{F}$ for $E<\Delta_0$. These approximations are also employed in $\eta_{s1}$ and $\eta_{s2}$ in the wave function basis $\psi_{e\uparrow(h\downarrow)}^{\pm}$ which contain the $x$ components of the wave vectors, i.e., $k_{x}^{e\uparrow}$ and $k_{x}^{h\downarrow}$.

For the wave functions in Eqs.(\ref{kdL})-(\ref{kdR}) solved for the injection of the $K\downarrow$ electrons, they can be derived from the BdG equations under the same approximations in a similarly way.

\end{document}